\documentclass[12pt]{article}

\usepackage{amssymb}

\usepackage[dvips]{graphicx}

\begin{document}

\title
{Factorization of integrals defining the two-loop $\beta$-function
for the general renormalizable $N=1$ SYM theory, regularized by
the higher covariant derivatives, into integrals of double total
derivatives}

\author{K.V.Stepanyantz}

\maketitle

\begin{center}

{\em Moscow State University, Physical Faculty,\\
Department of Theoretical Physics.\\
$119991$, Moscow, Russia}

\end{center}

\begin{abstract}
The integrals defining the two-loop $\beta$-function for the
general renormalizable $N=1$ supersymmetric Yang--Mills theory,
regularized by higher covariant derivatives, are investigated. It
is shown that they are given by integrals of double total
derivatives. These integrals are not equal to zero due to
appearing of $\delta$-functions. These $\delta$-functions allow to
reduce the two-loop integrals to one-loop integrals, which can be
easily calculated. The result agrees with the exact NSVZ
$\beta$-function and calculations made by different methods.
\end{abstract}


\section{Introduction.}
\hspace{\parindent}

Quantum correction in supersymmetric theories were studied for a
long time. For example, the $\beta$-function for $N=1$
supersymmetric Yang--Mills theories was calculated in one-
\cite{1Loop}, two- \cite{2Loop}, three- \cite{N124_3Loop,Jones},
and four-loop \cite{Harlander} approximations. All these
calculation were made with the dimensional reduction \cite{Siegel}
in the $\overline{MS}$-scheme \cite{MS}, because the dimensional
regularization \cite{tHooft} breaks the supersymmetry. However, it
is well known \cite{Siegel2} that the dimensional reduction is
inconsistent. Although ways allowing to overcome the corresponding
problems are discussed in the literature \cite{Stockinger},
removing of inconsistencies leads to the loss of the supersymmetry
in higher orders \cite{Avdeev,AvdeevVladimirov}. In particular
\cite{Avdeev,Velizhanin}, it was shown that obtaining a three-loop
$\beta$-function by different methods (from different vertexes)
leads to different results for the $N=2$ supersymmetric
Yang--Mills theory. As a consequence, the dimensional reduction
scheme breaks the supersymmetry in higher loops. (In \cite{Avdeev}
it was argued that this also takes place for the $N=1$ and $N=4$
supersymmetric Yang--Mills theories in the three-loop
approximation, but recent calculation \cite{Velizhanin} showed
that in the three-loop approximation this is true only for the
$N=2$ theory. For the $N=4$ supersymmetric Yang--Mills theory the
dimensional reduction does not also break the supersymmetry in the
four-loop approximation \cite{N4_4Loop}.)

Other regularizations are also applied to calculations of quantum
corrections. For example, in Ref. \cite{OperatorExpansion} the
two-loop $\beta$-function of $N=1$ supersymmetric electrodynamics
(and also the $\beta$-functions of the scalar and spinor
electrodynamics) was calculated using a method based on the
operator product expansion. The two-loop $\beta$-function for
$N=1$ supersymmetric Yang--Mills theory was calculated in Ref.
\cite{Mas} with the differential renormalization \cite{DiffR}.
Some calculations were made with the higher covariant derivative
regularization, proposed in \cite{Slavnov}, and generalized to the
supersymmetric case in \cite{Krivoshchekov} (another variant was
proposed in \cite{West_Paper}). The higher covariant derivative
regularization is an invariant regularization and does not break
the supersymmetry \cite{Krivoshchekov,West_Paper,Jones_Reg}.
However, it was not frequently applied to concrete calculations,
because it is very difficult to calculate the corresponding loop
integrals analytically. For example, the one-loop $\beta$-function
of the (non-supersymmetric) Yang-Mills theory was first calculated
only in \cite{Ruiz}. Taking into account correction made in
subsequent papers \cite{Asorey} the result coincided with the
well-known one, obtained with the dimensional regularization
\cite{Politzer}. It is possible to prove that in the one-loop
approximation the results obtained with the higher derivative
regularization always agree with the results obtained with the
dimensional regularization \cite{PhysLett}. Some calculations in
the one-loop and two-loop approximations were made for various
theories \cite{Rosten_Calc,Rosten_Int} with a variant of the
higher covariant derivative regularization proposed in
\cite{Rosten_Reg}. The structure of the corresponding integrals
was discussed in Ref. \cite{Rosten_Int}.

Calculations of quantum corrections in supersymmetric theories
with the higher derivative regularization show that the
$\beta$-function is given by integrals of total derivatives. This
was first noted in \cite{3LoopHEP}, where all integrals defining
the three-loop $\beta$-function of $N=1$ supersymmetric
electrodynamics were calculated using integration by parts. This
feature was also found in Ref. \cite{Smilga}, where the
factorization of integrands into total derivatives is explained
using a special technique, based on the covariant Feynman rules in
the background field method \cite{Grisaru,Milewski}. A proof of
the factorization for $N=1$ SQED by a different method
\cite{Identity,FactorizationHEP} is made in \cite{Recent_HEP}.
This factorization allows natural explaining the origin of the
exact NSVZ $\beta$-function \cite{NSVZ}, because one of the loop
integrals can be calculated explicitly. As a consequence, say, in
$N=1$ SQED integrals defining the $\beta$-function in the $n$-th
loop are reduced to integrals defining the anomalous dimension in
the $(n-1)$-th loop \cite{NSVZ_SQED}. It is important to note that
with the higher derivative regularization in order to obtain the
NSVZ $\beta$-function one should not make a special redefinition
of the coupling constant \cite{Recent_HEP}, which is needed if the
calculations are made with the dimensional reduction
\cite{Jones,Four-Loop_Indirect}.

For the general renormalizable $N=1$ supersymmetric Yang--Mills
theory, regularized by the higher covariant derivatives, the
two-loop $\beta$-function have been calculated in
\cite{PhysLett2}. Similar results were obtained with two different
versions of the higher derivative regularization in
\cite{MIAN,ECHAYA}. In these papers it was also verified that all
integrals defining the $\beta$-function are integrals of total
derivatives, and this feature does not depend on a particular
choice of the regularizing term. However, in Ref. \cite{Smilga} it
was argued that the integrals defining the $\beta$-function are
integrals of double total derivatives. For $N=1$ SQED, regularized
by higher derivatives, this was also proved by a different method
in \cite{Recent_HEP}. In the present paper we demonstrate that for
a general renormalizable $N=1$ supersymmetric Yang--Mills theory,
regularized by the higher covariant derivatives, two-loop
integrals for the $\beta$-function can be also written as
integrals of double total derivatives.

The paper is organized as follows:

In Sec. \ref{Section_SUSY} we introduce the notation and recall
basic information about the higher covariant derivative
regularization. The integrals defining the $\beta$-function for
the considered theory are rewritten as integrals of double total
derivatives in Sec. \ref{Section_Matter_Contribution}. The result
is briefly discussed in the Conclusion.


\section{$N=1$ supersymmetric Yang--Mills theory and the higher covariant
derivative regularization}
\hspace{\parindent}\label{Section_SUSY}

In this paper we consider a general renormalizable $N=1$
supersymmetric Yang--Mills theory. In the massless case it is
described by the action \cite{West,Buchbinder}\footnote{In our
notation $\eta_{\mu\nu}=\mbox{diag}(1,-1,-1,-1)$;\ \ $\theta^a
\equiv \theta_b C^{ba}$; $\theta_a$ and $\bar\theta_a$ denote the
right and left components of $\theta$, respectively.}

\begin{eqnarray}\label{SYM_Action}
&& S = \frac{1}{2 e^2} \mbox{Re}\,\mbox{tr}\int
d^4x\,d^2\theta\,W_a C^{ab} W_b + \frac{1}{4}\int d^4x\,
d^4\theta\, (\phi^*)^i (e^{2V})_i{}^j\phi_j +\nonumber\\
&&\qquad\qquad\qquad\qquad\qquad\qquad\qquad\qquad +
\Bigg(\frac{1}{6} \int d^4x\, d^2\theta\,\lambda^{ijk} \phi_i
\phi_j \phi_k + \mbox{h.c.}\Bigg),\qquad
\end{eqnarray}

\noindent where $\phi_i$ are chiral matter superfields in a
representation $R$, which is in general reducible. $V$ is a real
scalar gauge superfield. The superfield $W_a$ is a supersymmetric
gauge field stress tensor, which is defined by

\begin{equation}
W_a = \frac{1}{8} \bar D^2 (e^{-2V} D_a e^{2V}).
\end{equation}

\noindent In our notation $D_a$ and $\bar D_a$ are the right and
left supersymmetric covariant derivatives respectively, $V = e V^A
T^A$, and the generators of the fundamental representation are
normalized by the condition

\begin{equation}
\mbox{tr}(t^A t^B) = \frac{1}{2}\delta^{AB}.
\end{equation}

\noindent Because action (\ref{SYM_Action}) should be invariant
under the gauge transformations, the coefficient $\lambda^{ijk}$
satisfies the condition

\begin{equation}\label{Lambda_Invariance}
(T^A)_m{}^{i}\lambda^{mjk} + (T^A)_m{}^{j}\lambda^{imk} +
(T^A)_m{}^{k}\lambda^{ijm} = 0.
\end{equation}

It is convenient to calculate quantum corrections using the
background field method \cite{West}. We make the substitution

\begin{equation}\label{Substitution}
e^{2V} \to e^{2V'} \equiv e^{\mbox{\boldmath${\scriptstyle
\Omega}$}^+} e^{2V} e^{\mbox{\boldmath${\scriptstyle \Omega}$}}
\end{equation}

\noindent in action (\ref{SYM_Action}), where
$\mbox{\boldmath${\Omega}$}$ is a background superfield. Then the
theory is invariant under the background gauge transformations

\begin{equation}\label{Background_Transformations}
\phi \to e^{i\Lambda}\phi;\quad  V \to e^{iK} V e^{-iK}; \quad
e^{\mbox{\boldmath${\scriptstyle \Omega}$}} \to e^{iK}
e^{\mbox{\boldmath${\scriptstyle \Omega}$}} e^{-i\Lambda};\quad
e^{\mbox{\boldmath${\scriptstyle \Omega}$}^+} \to e^{i\Lambda^+}
e^{\mbox{\boldmath${\scriptstyle \Omega}$}^+} e^{-iK},
\end{equation}

\noindent where $K$ is an arbitrary real superfield, and $\Lambda$
is a background-chiral superfield. This invariance allows to set
$\mbox{\boldmath$\Omega$} = \mbox{\boldmath$\Omega$}^+ = {\bf V}$.
We choose a regularization and a gauge fixing so that invariance
(\ref{Background_Transformations}) is unbroken. A gauge is fixed
by adding

\begin{equation}\label{Gauge_Fixing}
S_{\mbox{\scriptsize gf}} = - \frac{1}{32 e^2}\,\mbox{tr}\,\int
d^4x\,d^4\theta\, \Big(V \mbox{\boldmath$D$}^2
\bar{\mbox{\boldmath$D$}}^2  V + V \bar {\mbox{\boldmath$D$}}^2
\mbox{\boldmath$D$}^2 V\Big)
\end{equation}

\noindent to the action. The actions for the corresponding
Faddeev--Popov and Nielsen--Kallosh ghosts are

\begin{eqnarray}
&& S_{\mbox{\scriptsize FP}} = \frac{1}{2e^2} \mbox{tr}\int
d^4x\,d^4\theta\, (e^{\mbox{\boldmath${\scriptstyle \Omega}$}}
\tilde c e^{-\mbox{\boldmath${\scriptstyle \Omega}$}} +
e^{-\mbox{\boldmath${\scriptstyle \Omega}$}^+} \tilde c^+
e^{\mbox{\boldmath${\scriptstyle \Omega}$}^+})\Big(
V_{\mbox{\scriptsize Ad}} (e^{\mbox{\boldmath${\scriptstyle
\Omega}$}} c e^{-\mbox{\boldmath${\scriptstyle \Omega}$}} +
e^{-\mbox{\boldmath${\scriptstyle \Omega}$}^+} c^+
e^{\mbox{\boldmath${\scriptstyle \Omega}$}^+})\nonumber\\
&& + V_{\mbox{\scriptsize Ad}} \mbox{cth}\, V_{\mbox{\scriptsize
Ad}} (e^{\mbox{\boldmath${\scriptstyle \Omega}$}} c
e^{-\mbox{\boldmath${\scriptstyle \Omega}$}} -
e^{-\mbox{\boldmath${\scriptstyle \Omega}$}^+} c^+
e^{\mbox{\boldmath${\scriptstyle \Omega}$}^+})\Big);\nonumber\\
&& S_{\mbox{\scriptsize NK}} = \frac{1}{2e^2} \mbox{tr}\int
d^4x\,d^4\theta\,b^+ e^{\mbox{\boldmath${\scriptstyle \Omega}$}^+}
e^{\mbox{\boldmath${\scriptstyle \Omega}$}} b\,
e^{-\mbox{\boldmath${\scriptstyle \Omega}$}}
e^{-\mbox{\boldmath${\scriptstyle \Omega}$}^+},
\end{eqnarray}

\noindent where

\begin{equation}
f(V_{\mbox{\scriptsize Ad}}) c = f(0) c + \frac{1}{1!} f'(0)[V,c]
+ \frac{1}{2!} f''(0) [V,[V,c]] + \ldots
\end{equation}

In order to introduce the regularization it is necessary to add
terms with the higher covariant derivatives to the action. There
are different possibilities for choosing such terms. For example,
in \cite{PhysLett2} the following terms were added:

\begin{eqnarray}\label{Regularized_Action1}
&& S_\Lambda = \frac{1}{2 e^2}\mbox{tr}\,\mbox{Re}\int
d^4x\,d^4\theta\, V\frac{(\mbox{\boldmath$D$}_\mu^2)^{n+1}}{
\Lambda^{2n}} V + \frac{1}{8} \int
d^4x\,d^4\theta\,\Bigg((\phi^*)^i
\Big[e^{\mbox{\boldmath${\scriptstyle \Omega}$}^+} e^{2V}
\frac{(\mbox{\boldmath$D$}_\alpha^2)^{m}}{\Lambda^{2m}}
e^{\mbox{\boldmath${\scriptstyle \Omega}$}}\Big]{}_i{}^j \phi_j
+\nonumber\\
&& + (\phi^*)^i \Big[e^{\mbox{\boldmath${\scriptstyle \Omega}$}^+}
\frac{(\mbox{\boldmath$D$}_\alpha^2)^{m}}{\Lambda^{2m}} e^{2V}
e^{\mbox{\boldmath${\scriptstyle \Omega}$}}\Big]{}_i{}^j
\phi_j\Bigg),
\end{eqnarray}

\noindent where $\mbox{\boldmath$D$}_\alpha$ is the background
covariant derivative and we assume that $m<n$. Below we call this
choice "variant 1". It is important that the higher covariant
derivative term is also introduced for the matter superfields,
because the considered theory contains a nontrivial
superpotential.

A simpler variant of the regularization is obtained if terms with
the higher covariant derivatives are chosen in the form ("variant
2")\cite{MIAN}

\begin{equation}\label{Regularized_Action2}
S_\Lambda = \frac{1}{2 e^2}\mbox{tr}\,\mbox{Re}\int
d^4x\,d^4\theta\, V\frac{(\mbox{\boldmath$D$}_\mu^2)^{n+1}}{
\Lambda^{2n}} V + \frac{1}{4} \int d^4x\,d^4\theta\,(\phi^*)^i
\Big[e^{\mbox{\boldmath${\scriptstyle \Omega}$}^+}
\frac{(\mbox{\boldmath$D$}_\alpha^2)^{m}}{\Lambda^{2m}}
e^{\mbox{\boldmath${\scriptstyle \Omega}$}}\Big]{}_i{}^j \phi_j.
\end{equation}

\noindent where $m$ and $n$ are arbitrary positive integers.

In both cases the regularized theory is evidently invariant under
the background gauge transformations. However, the higher
derivative terms considered here break BRST-invariance of the
action, and it is necessary to use a special subtraction scheme,
which restore the Slavnov--Taylor identities in each order of the
perturbation theory \cite{Slavnov12}. For the supersymmetric case
such a scheme was constructed in Ref. \cite{Slavnov34}.

After adding $S_\Lambda$ divergences remain only in the one-loop
approximation \cite{Slavnov_Book}. In order to regularize them, it
is necessary to introduce into the generating functional the
Pauli--Villars determinants

\begin{equation}
\prod\limits_{I}\Big(\int D\phi_{I}^* D\phi_{I}
e^{iS_{I}}\Big)^{-c_I} \prod\limits_{i}\Big(\int D c_{i}^+ D c_{i}
D \tilde c_{i}^+ D \tilde c_{i} Db_{i}^+ Db_i
e^{iS_{i}}\Big)^{-c_i},
\end{equation}

\noindent where $S_I$ and $S_i$ are the actions for the
Pauli--Villars fields corresponding to $\phi$ and ghosts,
respectively. For variant 1 (if $S_\Lambda$ is given by Eq.
(\ref{Regularized_Action1})), the Pauli--Villars action can be
chosen as \cite{PhysLett}

\begin{eqnarray}\label{PV1}
&& S_I = \frac{1}{8} \int d^4x\,d^4\theta\,\Bigg((\phi_I^*)^i
\Big[e^{\mbox{\boldmath${\scriptstyle \Omega}$}^+} e^{2V}
\Big(1+\frac{(\mbox{\boldmath$D$}_\alpha^2)^{m}}{\Lambda^{2m}}\Big)
e^{\mbox{\boldmath${\scriptstyle \Omega}$}}\Big]{}_i{}^j
(\phi_I)_j + (\phi_I^*)^i \Big[e^{\mbox{\boldmath${\scriptstyle
\Omega}$}^+}
\Big(1+\frac{(\mbox{\boldmath$D$}_\alpha^2)^{m}}{\Lambda^{2m}}
\Big)\times\nonumber\\
&& \times e^{2V} e^{\mbox{\boldmath${\scriptstyle
\Omega}$}}\Big]{}_i{}^j (\phi_I)_j\Bigg) + \Big(\frac{1}{4}\int
d^4x\,d^2\theta\,M_I^{ij} (\phi_{I})_i (\phi_{I})_j +\mbox{h.c.}
\Big).
\end{eqnarray}

\noindent For variant 2 the Pauli--Villars action is

\begin{equation}\label{PV2}
S_I = \frac{1}{4} \int d^4x\,d^4\theta\,(\phi_I^*)^i
\Big[e^{\mbox{\boldmath${\scriptstyle \Omega}$}^+}
\Big(1+\frac{(\mbox{\boldmath$D$}_\alpha^2)^{m}}{\Lambda^{2m}}\Big)
e^{\mbox{\boldmath${\scriptstyle \Omega}$}}\Big]{}_i{}^j
(\phi_I)_j + \Big(\frac{1}{4}\int d^4x\,d^2\theta\,M_I^{ij}
(\phi_{I})_i (\phi_{I})_j +\mbox{h.c.} \Big).
\end{equation}

\noindent The mass terms for the ghost Pauli--Villars fields are

\begin{eqnarray}
\frac{1}{2e^2}\mbox{tr}\int d^4x\,d^2\theta\,\Big(m_b b^2 + 2 m_c
\tilde c c \Big)+ \mbox{h.c.}
\end{eqnarray}

\noindent The masses of all Pauli--Villars fields are proportional
to the parameter $\Lambda$:

\begin{equation}
M^{ij}_I = a_I^{ij}\Lambda;\qquad m_i = a_i\Lambda,
\end{equation}

\noindent where $a$-s are numerical constants. As a consequence,
$\Lambda$ is the only dimensionful parameter of the regularized
theory. We assume that the mass term does not break the gauge
invariance. Also we will choose the masses so that

\begin{equation}
M_I^{ij} (M_I^*)_{jk} = M_I^2 \delta_k^i.
\end{equation}

\noindent The coefficients $c_I$ and $c_i$ satisfy the conditions

\begin{equation}
\sum\limits_I c_I = 1;\qquad \sum\limits_I c_I M_I^2 = 0;\qquad
\sum\limits_i c_i = 1;\qquad \sum\limits_i c_i m_i^2 = 0.
\end{equation}

The generating functional for connected Green functions and the
effective action are defined by the standard way.


\section{Two-loop $\beta$-function}
\label{Section_Matter_Contribution} \hspace{\parindent}

Let us write terms in the effective action corresponding to the
renormalized two-point Green function of the gauge superfield in
the form

\begin{equation}\label{D_Definition}
\Gamma^{(2)}_V = - \frac{1}{8\pi} \mbox{tr}\int
\frac{d^4p}{(2\pi)^4}\,d^4\theta\,{\bf V}(-p)\,\partial^2\Pi_{1/2}
{\bf V}(p)\, d^{-1}(\alpha,\lambda,\mu/p),
\end{equation}

\noindent where $\alpha$ is a renormalized coupling constant. In
this paper we investigate the expression

\begin{equation}
\frac{d}{d\ln \Lambda}\,
\Big(d^{-1}(\alpha_0,\lambda_0,\Lambda/p)-\alpha_0^{-1}\Big)\Big|_{p=0}
= - \frac{d\alpha_0^{-1}}{d\ln\Lambda} =
\frac{\beta(\alpha_0,\lambda_0)}{\alpha_0^2}
\end{equation}

\noindent in the two-loop approximation. After the calculation of
the supergraphs the two-loop $\beta$-function can be presented in
the form:

\begin{eqnarray}\label{Result}
&& \beta_2(\alpha,\lambda) = \alpha^2 C_2 (I_{\mbox{\scriptsize
FP}} + I_{\mbox{\scriptsize NK}}) + \alpha^2 T(R) I_0 + \alpha^3
C_2^2 I_1 + \frac{\alpha^3}{r} C(R)_i{}^j
C(R)_j{}^i I_2  +\quad\nonumber\\
&& + \alpha^3 T(R) C_2 I_3 + \alpha^2 C(R)_i{}^j
\frac{\lambda_{jkl}^* \lambda^{ikl}}{4\pi r} I_4,
\end{eqnarray}

\noindent where the following notation is used:

\begin{eqnarray}\label{T(R)}
&& \mbox{tr}\,(T^A T^B) \equiv T(R)\,\delta^{AB};\qquad
(T^A)_i{}^k
(T^A)_k{}^j \equiv C(R)_i{}^j;\nonumber\\
&& f^{ACD} f^{BCD} \equiv C_2 \delta^{AB};\qquad\quad r\equiv
\delta_{AA}.
\end{eqnarray}

\noindent Here

\begin{eqnarray}
&& I = I(0) -\sum\limits_{I} c_I I(M_I)\quad\mbox{for}\quad I_0,
I_2, I_3;\nonumber\\
&& I = I(0) -\sum\limits_{i} c_i I(m_i)\quad\mbox{for}\quad
I_{\mbox{\scriptsize NK}}, I_{\mbox{\scriptsize FP}},
\end{eqnarray}

\noindent and the integrals $I_0(M)$, $I_1$, $I_2(M)$, $I_3(M)$
and $I_4$ can be found in Ref. \cite{PhysLett2} for variant 1 and
in Ref. \cite{MIAN} for variant 2. In Refs. \cite{PhysLett2,MIAN}
these integrals are written as integrals of total derivatives.
However, they are actually the integrals of double total
derivatives. Note that in this paper the notation is different
from Ref. \cite{Recent_HEP}, where

\begin{equation}
\int \frac{d^4q}{(2\pi)^4} \frac{\partial}{\partial q^\mu} \equiv
\int\limits_{S_\infty} \frac{dS_\mu}{(2\pi)^4}
\end{equation}

\noindent corresponds to $\mbox{Tr} [x_\mu, \ldots]$. Here we use
the ordinary notation

\begin{equation}
\int \frac{d^4q}{(2\pi)^4} \frac{\partial}{\partial q^\mu} \equiv
\int\limits_{\partial} \frac{dS_\mu}{(2\pi)^4} =
\int\limits_{S_\infty} \frac{dS_\mu}{(2\pi)^4} - \mbox{integrals
of $\delta$-singularities},
\end{equation}

\noindent and $\partial$ denotes a boundary of a region where the
integrand is regular.

The result for the integrals defining the $\beta$-function for the
variant 1 can be written as follows (a two-loop contribution of
the Faddeev--Popov ghosts is 0, exactly as in \cite{Grisaru}):

\begin{eqnarray}
&& I_{\mbox{\scriptsize NK}}(m) = \frac{1}{2} I_{\mbox{\scriptsize
FP}}(m) = \pi \int \frac{d^4q}{(2\pi)^4} \frac{d}{d\ln\Lambda}
\frac{\partial}{\partial q^\mu} \frac{\partial}{\partial q_\mu}
\Bigg\{\frac{1}{q^2}\ln\Big(q^2+m^2\Big)\Bigg\};\\
&&\vphantom{1}\nonumber\\
&& I_0(M) = -\pi \int \frac{d^4q}{(2\pi)^4} \frac{d}{d\ln\Lambda}
\frac{\partial}{\partial q^\mu} \frac{\partial}{\partial q_\mu}
\Bigg\{\frac{1}{q^2}
\ln\Big(q^2(1+q^{2m}/\Lambda^{2m})^2+M^2\Big) \Bigg\};\\
&&\vphantom{1}\nonumber\\
&& I_1 = -12\pi^2 \int \frac{d^4q}{(2\pi)^4} \frac{d^4k}{(2\pi)^4}
\frac{d}{d\ln\Lambda} \frac{\partial}{\partial k^\mu}
\frac{\partial}{\partial k_\mu} \Bigg\{\frac{1}{k^2
(1+k^{2n}/\Lambda^{2n}) q^2 (1+q^{2n}/\Lambda^{2n})
(q+k)^2}\nonumber\\
&& \times \frac{1}{(1+(q+k)^{2n}/\Lambda^{2n})}  \Bigg\};\\
&&\vphantom{1}\nonumber\\
&& I_2(M) = 2\pi^2 \int \frac{d^4q}{(2\pi)^4}
\frac{d^4k}{(2\pi)^4} \frac{d}{d\ln\Lambda}
\frac{\partial}{\partial q^\mu} \frac{\partial}{\partial q_\mu}
\Bigg\{\frac{(2+(q+k)^{2m}/\Lambda^{2m}+ q^{2m}/\Lambda^{2m})^2
}{k^2 (1+k^{2n}/\Lambda^{2n})}
\nonumber\\
&& \times \frac{(1+ q^{2m}/\Lambda^{2m})(1+
(q+k)^{2m}/\Lambda^{2m})}{\Big(q^2(1+q^{2m}/\Lambda^{2m})^2
+M^2\Big)\Big((q+k)^2(1+ (q+k)^{2m}/\Lambda^{2m})^2 +M^2\Big)}
\Bigg\};\\
&&\vphantom{1}\nonumber\\
&& I_3(M) = 2\pi^2 \int \frac{d^4q}{(2\pi)^4}
\frac{d^4k}{(2\pi)^4} \frac{d}{d\ln\Lambda}
\frac{\partial}{\partial q^\mu} \frac{\partial}{\partial k_\mu}
\Bigg\{ \frac{(2+ k^{2m}/\Lambda^{2m} +
q^{2m}/\Lambda^{2m})^2}{(k+q)^2 (1+
(q+k)^{2n}/\Lambda^{2n})} \nonumber\\
&& \times
\frac{(1+k^{2m}/\Lambda^{2m})(1+q^{2m}/\Lambda^{2m})}{\Big(k^2(1+
k^{2m}/\Lambda^{2m})^2+M^2\Big)\Big(q^2(1+
q^{2m}/\Lambda^{2m})^2+M^2\Big)} \Bigg\};\\
&&\vphantom{1}\nonumber\\
&& I_4 = -8\pi^2 \int \frac{d^4q}{(2\pi)^4} \frac{d^4k}{(2\pi)^4}
\frac{d}{d\ln\Lambda} \frac{\partial}{\partial q^\mu}
\frac{\partial}{\partial q_\mu} \Bigg\{\frac{1}{k^2
(1+k^{2m}/\Lambda^{2m}) q^2 (1+q^{2m}/\Lambda^{2m}) (q+k)^2
}\nonumber\\
&& \times \frac{1}{(1+(q+k)^{2m}/\Lambda^{2m})} \Bigg\}.
\end{eqnarray}

\noindent These integrals are not equal to 0 because

\begin{equation}
\int \frac{d^4q}{(2\pi)^4} \frac{\partial}{\partial q^\mu}
\frac{\partial}{\partial q_\mu} \Big(\frac{f(q^2)}{q^2}\Big) =
\lim\limits_{\varepsilon\to 0}\int\limits_{S_\varepsilon}
\frac{dS_\mu}{(2\pi)^4} \frac{(-2) q^\mu f(q^2)}{q^4} =
\frac{1}{4\pi^2} f(0)
\end{equation}

\noindent for a nonsingular function $f(q^2)$ which rapidly
decreases at the infinity. As a consequence,

\begin{eqnarray}
&& I_{\mbox{\scriptsize NK}} = -\frac{1}{4\pi}
\frac{d}{d\ln\Lambda}\Big(\sum\limits_i
c_i \ln m_i^2 \Big) = -\frac{1}{2\pi};\nonumber\\
&& I_0 = \frac{1}{4\pi} \frac{d}{d\ln\Lambda}\Big(\sum\limits_I
c_I \ln M_I^2 \Big) = \frac{1}{2\pi};\nonumber\\
&& I_1 = - 6 \int \frac{d^4q}{(2\pi)^4} \frac{d}{d\ln\Lambda}
\Bigg[\frac{1}{q^4 (1+q^{2n}/\Lambda^{2n})^2} \Bigg] =
- \frac{3}{4\pi^2};\nonumber\\
&& I_2 = \int \frac{d^4k}{(2\pi)^4} \frac{d}{d\ln\Lambda}
\Bigg[\frac{(2+k^{2m}/\Lambda^{2m})^2}{k^4
(1+k^{2n}/\Lambda^{2n})(1+k^{2m}/\Lambda^{2m})} \Bigg] =
\frac{1}{2\pi^2};\nonumber\\
&& I_3 = \int \frac{d^4q}{(2\pi)^4} \frac{d}{d\ln\Lambda}
\Bigg[\frac{2}{q^4} -\sum\limits_I c_I
\frac{2(1+q^{2m}/\Lambda^{2m})^4}{(q^2(1+q^{2m}/\Lambda^{2m})^2
+ M_I^2)^2} \Bigg] = \frac{1}{4\pi^2};\nonumber\\
&& I_4 = - 4\int \frac{d^4k}{(2\pi)^4} \frac{d}{d\ln\Lambda}
\Bigg[\frac{1}{k^4 (1+k^{2m}/\Lambda^{2m})^2} \Bigg] =
-\frac{1}{2\pi^2}.
\end{eqnarray}

\noindent (The Pauli--Villars fields nontrivially contribute only
to integrals $I_{\mbox{\scriptsize NK}}$, $I_{\mbox{\scriptsize
FP}}$, $I_0$ and $I_3$, where they cancel the one-loop
(sub)divergence.)

For variant 2 the integrals $I_{\mbox{\scriptsize FP}}$,
$I_{\mbox{\scriptsize NK}}$, $I_0$, $I_1$, and $I_4$ are the same.
However, the integrals $I_2$ and $I_3$ are different:

\begin{eqnarray}
&& I_2(M) = 8\pi^2 \int \frac{d^4q}{(2\pi)^4}
\frac{d^4k}{(2\pi)^4} \frac{d}{d\ln\Lambda}
\frac{\partial}{\partial q^\mu} \frac{\partial}{\partial q_\mu}
\Bigg\{\frac{1}{k^2 (1+k^{2n}/\Lambda^{2n})}
\nonumber\\
&& \times \frac{(1+ q^{2m}/\Lambda^{2m})(1+
(q+k)^{2m}/\Lambda^{2m})}{\Big(q^2(1+q^{2m}/\Lambda^{2m})^2
+M^2\Big)\Big((q+k)^2(1+ (q+k)^{2m}/\Lambda^{2m})^2 +M^2\Big)}
\Bigg\};\\
&&\vphantom{1}\nonumber\\
&& I_3(M) = 8\pi^2 \int \frac{d^4q}{(2\pi)^4}
\frac{d^4k}{(2\pi)^4} \frac{d}{d\ln\Lambda}
\frac{\partial}{\partial q^\mu} \frac{\partial}{\partial k_\mu}
\Bigg\{\frac{1}{(k+q)^2 (1+
(q+k)^{2n}/\Lambda^{2n})} \nonumber\\
&& \times
\frac{(1+k^{2m}/\Lambda^{2m})(1+q^{2m}/\Lambda^{2m})}{\Big(k^2(1+
k^{2m}/\Lambda^{2m})^2+M^2\Big)\Big(q^2(1+
q^{2m}/\Lambda^{2m})^2+M^2\Big)} \Bigg\}.
\end{eqnarray}

\noindent As a consequence,

\begin{eqnarray}
&& I_2 = \int \frac{d^4k}{(2\pi)^4} \frac{d}{d\ln\Lambda}
\Bigg[\frac{4}{k^4 (1+k^{2n}/\Lambda^{2n})(1+k^{2m}/\Lambda^{2m})}
\Bigg] =\frac{1}{2\pi^2};\\
&& I_3 = \int \frac{d^4q}{(2\pi)^4} \frac{d}{d\ln\Lambda}
\Bigg[\frac{2}{q^4 (1+q^{2m}/\Lambda^{2m})^2} -\sum\limits_I c_I
\frac{2(1+q^{2m}/\Lambda^{2m})^2}{(q^2(1+q^{2m}/\Lambda^{2m})^2 +
M_I^2)^2} \Bigg] = \frac{1}{4\pi^2}.\nonumber
\end{eqnarray}

\noindent Therefore, for both variants of the regularization the
two-loop $\beta$-function is given by

\begin{eqnarray}
&& \beta(\alpha,\lambda) = - \frac{\alpha^2}{2\pi}\Big(3 C_2 -
T(R)\Big) + \frac{\alpha^3}{(2\pi)^2}\Big(-3 C_2^2 + T(R) C_2 +
\frac{2}{r}
C(R)_i{}^j C(R)_j{}^i\Big) -\nonumber\\
&& - \frac{\alpha^2 C(R)_i{}^j \lambda_{jkl}^*
\lambda^{ikl}}{8\pi^3 r} + \ldots
\end{eqnarray}

\noindent and agrees with the exact NSVZ $\beta$-function
\cite{NSVZ,NSVZ_SQED}

\begin{equation}\label{NSVZ_Beta}
\beta(\alpha,\lambda) = - \frac{\alpha^2\Big[3 C_2 - T(R) +
C(R)_i{}^j \gamma_j{}^i(\alpha,\lambda)/r \Big]}{2\pi(1-
C_2\alpha/2\pi)}.
\end{equation}

\noindent Up to notation, this result is in agreement with the
results of calculations made with the dimensional reduction in
\cite{2Loop}.


\section{Conclusion}
\label{Section_Conclusion} \hspace{\parindent}

With the higher covariant derivative regularization all integrals
defining the two-loop $\beta$-function of the general
renormalizable $N=1$ supersymmetric Yang--Mills theory are
integrals of total derivatives. In this paper using two different
versions of the higher covariant derivative regularization we show
that they are not only integrals of total derivatives, but also
integrals of double total derivatives. Due to the identity

\begin{equation}
\frac{\partial}{\partial q^\mu} \frac{\partial}{\partial q_\mu}
\frac{1}{q^2} = - 4\pi^2 \delta^4(q)
\end{equation}

\noindent these integrals do not vanish. Calculating them one
obtain the exact NSVZ $\beta$-function. Possibly, this situation
also takes place in all orders of the perturbation theory.

\bigskip
\bigskip

\noindent {\Large\bf Acknowledgements.}

\bigskip

\noindent This work was partially supported by RFBR grant No
11-01-00296a. I am very grateful to prof. A.L.Kataev for valuable
discussions.



\begin{thebibliography}{100}


\bibitem{1Loop}
{\it S.Ferrara, B.Zumino}, Nucl.Phys. {\bf B79} (1974) 413.

\bibitem{2Loop}
{\it D.R.T.Jones}, Nucl.Phys. {\bf B87} (1975) 127.

\bibitem{N124_3Loop}
{\it L.V.Avdeev, O.V.Tarasov}, Phys.Lett. {\bf 112 B} (1982) 356.

\bibitem{Jones}
{\it I.Jack, D.R.T.Jones, C.G.North}, Phys.Lett {\bf B386} (1996)
138; Nucl.Phys. {\bf B 486} (1997) 479.

\bibitem{Harlander}
{\it R.V.Harlander, D.R.T.Jones, P.Kant, L.Mihaila,
M.Steinhauser}, JHEP {\bf 0612} (2006) 024.

\bibitem{Siegel}
{\it W.Siegel}, Phys.Lett. {\bf 84 B}, (1979), 193.

\bibitem{MS}
{\it W.A.Bardeen, A.J.Buras, D.W.Duke, and T.Muta}, Phys.Rev. {\bf
D18} (1978) 3998.

\bibitem{tHooft}
{\it G.t'Hooft, M.Veltman}, Nucl.Phys. {\bf B44}, (1972), 189.

\bibitem{Siegel2}
{\it W.Siegel}, Phys.Lett. {\bf 94B}, (1980), 37.

\bibitem{Stockinger}
{\it L.V.Avdeev, G.A.Chochia, A.A.Vladimirov}, Phys.Lett. {\bf
B105}, (1981), 272; {\it D.St\"{o}ckinger}, JHEP {\bf 0503},
(2005), 076.

\bibitem{Avdeev}
{\it L.V.Avdeev}, Phys.Lett. {\bf B117}, (1982), 317.

\bibitem{AvdeevVladimirov}
{\it L.V.Avdeev, A.A.Vladimirov}, Nucl.Phys. {\bf B219}, (1983),
262.

\bibitem{Velizhanin}
{\it V.N.Velizhanin}, Nucl.Phys. {\bf B818}, (2009), 95.

\bibitem{N4_4Loop}
{\it V.N. Velizhanin}, Phys.Lett. {\bf B696} (2011) 560.

\bibitem{OperatorExpansion}
{\it M.A.Shifman, A.I.Vainshtein} Sov.J.Nucl.Phys. {\bf 44},
(1986), 321.

\bibitem{Mas}
{\it J.Mas, M.Perez-Victoria, C.Seijas}, JHEP, {\bf 0203}, (2002),
049.

\bibitem{DiffR}
{\it D.Z.Freedman, K.Johnson, J.I.Latorre}, Nucl.Phys. {\bf B371},
(1992), 353.

\bibitem{Slavnov}
{\it A.A.Slavnov}, Nucl.Phys., {\bf B31}, (1971), 301;
Theor.Math.Phys. {\bf 13}, (1972), 1064.

\bibitem{Krivoshchekov}
{\it V.K.Krivoshchekov}, Theor.Math.Phys. {\bf 36}, (1978), 745.

\bibitem{West_Paper}
{\it P.West}, Nucl.Phys. {\bf B 268}, (1986), 113.

\bibitem{Jones_Reg}
{\it I.Jack, D.R.T.Jones}, Regularisation of supersymmetric
theories, hep-ph/9707278.

\bibitem{Ruiz}
{\it C.Martin, F.Ruiz Ruiz}, Nucl.Phys. {\bf B 436}, (1995), 545.

\bibitem{Asorey}
{\it M.Asorey, F.Falceto}, Phys.Rev {\bf D 54}, (1996), 5290;
{\it T.Bakeyev, A.Slavnov}, Mod.Phys.Lett. {\bf A11}, (1996),
1539.

\bibitem{Politzer}
{\it D.J.Gross, F.Wilczek}, Phys.Rev.Lett. {\bf 30}, (1973), 1343;
{\it H.D.Politzer}, Phys.Rev.Lett. {\bf 30}, (1973), 1346.

\bibitem{PhysLett}
{\it P.Pronin, K.Stepanyantz}, Phys.Lett. {\bf B414}, (1997), 117.

\bibitem{Rosten_Calc}
{\it S.~Arnone, T.~R.~Morris, O.~J.~Rosten}, JHEP {\bf 0510},
(2005), 115; {\it T.~R.~Morris, O.~J.~Rosten}, J.Phys. {\bf A39},
(2006), 11657; {\it O.~J.~Rosten}, On the renormalization of
theories of a scalar chiral superfield, arXiv:0808.2150 [hep-th].

\bibitem{Rosten_Int}
{\it S.~Arnone, A.~Gatti, T.~R.~Morris, O.~J.~Rosten}, Phys.Rev.
{\bf D69}, (2004), 065009; {\it T.~R.~Morris, O.~J.~Rosten},
Phys.Rev. {\bf D73}, (2006), 065003;

\bibitem{Rosten_Reg}
{\it S.Arnone, Y.A.Kubyshin, T.R.Morris, J.F.Tighe},
Int.J.Mod.Phys. {\bf A17}, (2002), 2283.

\bibitem{3LoopHEP}
{\it A.A.Soloshenko, K.V.Stepanyantz}, hep-th/0304083;
Theor.Math.Phys. {\bf 140}, (2004), 1264.

\bibitem{Smilga}
{\it A.Smilga, A.Vainshtein}, Nucl.Phys. {\bf B 704}, (2005), 445.

\bibitem{Grisaru}
{\it M.T.Grisaru, D.Zanon}, Nucl.Phys. {\bf B252} (1985) 578.

\bibitem{Milewski}
{\it M.T.Grisaru, B.Milewski, D.Zanon}, Nucl.Phys. {\bf B266}
(1986) 589.

\bibitem{Identity}
{\it K.Stepanyantz}, Theor.Math.Phys. {\bf 146} (2006) 321.

\bibitem{FactorizationHEP}
{\it K.V.Stepanyantz}, Factorization of integrals, defining the
$\beta$-function, into integrals of total derivatives in $N=1$
SQED, regularized by higher derivatives, ArXiv:1101.2956 [hep-th].

\bibitem{Recent_HEP}
{\it K.V.Stepanyantz}, Nucl.Phys. {\bf B 852}, (2011), 71.

\bibitem{NSVZ}
{\it V.Novikov, M.Shifman, A.Vainshtein, V.Zakharov}, Nucl.Phys.
{\bf B 229}, (1983), 381; Phys.Lett. {\bf 166B}, (1985), 329; {\it
Shifman M.A., Vainshtein A.I.}, Nucl.Phys. {\bf B 277}, (1986),
456; Sov.Phys.JETP {\bf 64}, (1986) 428.

\bibitem{NSVZ_SQED}
{\it A.I.Vainshtein, V.I.Zakharov, M.A.Shifman}, JETP Lett. {\bf
42} (1985) 224; {\it M.Shifman, A.Vainshtein, V.Zakharov}, Phys.
Lett. {\bf B166} (1986) 334.

\bibitem{Four-Loop_Indirect} {\it I.Jack, D.R.T.Jones,
A.Pickering}, Phys. Lett. {\bf B435} (1998) 61.

\bibitem{PhysLett2}
{\it A.B.Pimenov, E.S.Shevtsova, K.V.Stepanyantz}, Phys.Lett. {\bf
B 686}, (2010), 293.

\bibitem{MIAN}
{\it K.V.Stepanyantz}, Proceedings of the Steklov Institute of
Mathematics, {\bf 272}, (2011), 256.

\bibitem{ECHAYA}
{\it K.V.Stepanyantz}, Phys.Part.Nucl.Lett. {\bf 8}, (2011), 321.

\bibitem{West}
{\it P.West}, Introduction to supersymmetry and supergravity,
World Scientific, 1986.

\bibitem{Buchbinder}
{\it I.L.Buchbinder, S.M.Kuzenko}, {Ideas and methods of
supersymmetry and supergravity}, Bristol and Philadelphia,
Institute of Physics Publishing, 1998.

\bibitem{Slavnov12}
{\it A.A.Slavnov}, Phys.Lett. {\bf B 518}, (2001), 195;
Theor.Math.Phys. {\bf 130}, (2002), 1.

\bibitem{Slavnov34}
{\it A.A.Slavnov, K.V.Stepanyantz}, Theor.Math.Phys., {\bf 135},
(2003), 673; {\bf 139}, (2004), 599.

\bibitem{Slavnov_Book}
{\it L.D.Faddeev, A.A.Slavnov}, Gauge fields, introduction to
quantum theory, second edition, Benjamin, Reading, 1990.


\end{thebibliography}
\end{document}